\documentclass[12pt,preprint]{aastex}

\slugcomment{accepted to the {\it Astronomical Journal} 31 July 2006}

\shorttitle{Parallaxes from CTIOPI --- RECONS Stars}
\shortauthors{Henry et al.}

\begin{document}

\title{The Solar Neighborhood XVII: Parallax Results from the CTIOPI
0.9 m Program -- Twenty New Members of the RECONS 10 Parsec Sample}

\author{Todd J. Henry\altaffilmark{1}, Wei-Chun Jao\altaffilmark{1},
John P. Subasavage\altaffilmark{1}, and Thomas
D. Beaulieu\altaffilmark{1}}

\affil{Georgia State University, Atlanta, GA 30302-4106}
\email{thenry@chara.gsu.edu, jao@chara.gsu.edu, 
subasavage@chara.gsu.edu, beaulieu@chara.gsu.edu}

\author{Philip A. Ianna\altaffilmark{1}} \affil{University of
Virginia, Charlottesville, VA 22903} \email{pai@virginia.edu}

\and

\author{Edgardo Costa\altaffilmark{1}, Ren\'{e}
A. M\'{e}ndez\altaffilmark{1}} \affil{Universidad de Chile, Santiago,
Chile} \email{costa@das.uchile.cl, rmendez@das.uchile.cl}

\altaffiltext{1}{Visiting Astronomer, Cerro Tololo Inter-American
Observatory.  CTIO is operated by AURA, Inc.\ under contract to the
National Science Foundation.}

\begin{abstract}

Astrometric measurements for 25 red dwarf systems are presented,
including the first definitive trigonometric parallaxes for 20 systems
within 10 pc of the Sun, the horizon of the RECONS sample.  The three
nearest systems that had no previous trigonometric parallaxes (other
than perhaps rough preliminary efforts) are SO 0253+1652 (3.84 $\pm$
0.04 pc, the 23rd nearest system), SCR 1845-6357 AB (3.85 $\pm$ 0.02
pc, 24th), and LHS 1723 (5.32 $\pm$ 0.04 pc, 56th).  In total, seven
of the systems reported here rank among the nearest 100 stellar
systems.  Supporting photometric and spectroscopic observations have
been made to provide full characterization of the systems, including
complete $VRIJHK_s$ photometry and spectral types.  A study of the
variability of 27 targets reveals six obvious variable stars,
including GJ 1207, for which we observed a flare event in the $V$ band
that caused it to brighten by 1.7 mag.

Improved parallaxes for GJ 54 AB and GJ 1061, both important members
of the 10 pc sample, are also reported.  Definitive parallaxes for GJ
1001 A, GJ 633, and GJ 2130 ABC, all of which have been reported to be
within 10 pc, indicate that they are beyond 10 pc.  From the analysis
of systems with (previously) high trigonometric parallax errors, we
conclude that parallaxes with errors in excess of 10 mas are
insufficiently reliable for inclusion in the RECONS sample.  The
cumulative total of new additions to the 10 pc sample since 2000 is
now 34 systems --- 28 by the RECONS team and six by other groups.
This total represents a net increase of 16\% in the number of stellar
systems reliably known to be nearer than 10 pc.

\end{abstract}

\keywords{surveys --- astrometry --- stars: distances --- stars: low
mass, brown dwarfs --- stars: statistics --- solar neighborhood}

\section{Introduction}
\label{sec:intro}

Trigonometric parallax determinations provide one of the most
fundamental measures of the cosmos.  Their beauty lies in their
simplicity: they are based only on geometric techniques and are the
straightforward result of the Earth orbiting the Sun.  Trigonometric
parallaxes help define solar neighborhood membership, allow for
accurate masses calculations in binary systems, provide distances to
fundamental stellar clusters, and supply benchmarks for stellar
population studies in the Milky Way and nearby galaxies.  With
concerted effort, trigonometric parallaxes with accuracies of $\sim$1
milliarcsecond are possible, yielding distances good to 10\% even at
100 pc, and allow astronomers to create an accurate three-dimensional
map of the nearby Galaxy.  Much like early explorers' discoveries
created a more complete map of our Earth, accurate trigonometric
parallaxes create a more detailed picture of the solar neighborhood,
filling in the blank regions of nearby space.

\citet{1995yCat.1174....0V} compiled decades of ground-based work by
dedicated astrometrists in {\it The General Catalogue of Trigonometric
Stellar Parallaxes} (also known as the ``Yale Parallax Catalog'',
hereafter, YPC).  The YPC includes 15994 total trigonometric
parallaxes for 8112 stars measured using photographic plates and CCDs
published before the end of 1995.  The Hipparcos space mission,
launched in August 1989, provided $\sim$118,000 parallaxes \citep[
hereafter HIP]{1997A&A...323L..49P,1997yCat.1239....0E}, and was
essentially complete for stars with $V = 7.3-9.0$, while reaching to
observe a few dozen stars even fainter than $V$$\sim$13.  However,
Hipparcos could not measure most nearby white, red, and brown dwarfs,
which comprise more than 80\% of the solar neighborhood population,
and many of these systems have not yet had their distances measured
via ground-based trigonometric parallax efforts.

Here we present results for nearby star systems observed during our
Cerro Tololo Inter-american Observatory Parallax Investigation
(CTIOPI).  Twenty systems are new members of the Research Consortium
on Nearby Stars (RECONS) sample, meeting the RECONS requirements that
they have trigonometric parallaxes, $\pi_{trig}$, larger than 100 mas
with errors less than 10 mas (see section~\ref{sec:astr.reconsdefn}).
We provide a brief dossier for each new systems, including $VRIJHK_s$
photometry and spectral types.  We also provide improved $\pi_{trig}$
for two additional important systems previously known to be within 10
pc, GJ 54 AB and GJ 1061, and for three systems that have reported
$\pi_{trig}$ larger than 100 mas, but which, in truth, lie beyond the
10 pc horizon.  In total, we report 27 new trigonometric parallax
measurements for 25 systems.

\section{CTIOPI and Other Recent Trigonometric Parallax Efforts}
\label{sec:parallaxes}

Since the publication of the YPC and HIP results, only $\sim$250
ground-based parallaxes and a handful space-based parallaxes from
$HST$ \citep[e.g.,][]{1999AJ....118.1086B} have been published in the
refereed literature.  We provide a comprehensive tally of ground-based
trigonometric parallax measurements published for new RECONS systems
in Table 1, since the 1995 cutoff date of the YPC.\footnote{All
parallaxes in \citet{2000AJ....119..369R} were updated in
\citet{2002AJ....124.1170D}.  Two objects, TVLM 513-46546
\citep{1995AJ....110.3014T,2002AJ....124.1170D} and 2MA 1047+2124,
\citep{2003AJ....126..975T,2004AJ....127.2948V} are borderline objects
--- the Tinney parallaxes place the objects closer than 10 pc whereas
the Dahn and Vrba USNO parallaxes place them beyond 10 pc.  Weighted
means of the two values for each object yield distances larger than 10
pc, so neither is included as a new 10 pc system in Table 1.}  The
transition of data acquisition techniques from photographic plates to
CCDs and infrared arrays is nearly complete, although approximate
parallaxes for the nearest objects can still be derived from large
archival databases that include scanned plates
\citep{2001A&A...380..148D,2003ApJ...589L..51T,2005AJ....129..409D}.
These parallaxes are useful but necessarily crude, given the limited
number of epochs available and the coarse plate scales.  The most
recent change in parallax work is the advent of IR arrays
\citep[e.g.,][]{2004AJ....127.2948V}.

Our CCD parallax effort, CTIOPI, began as an NOAO Surveys Program in
1999 August using both the 0.9 m and 1.5 m telescopes at CTIO, and has
continued on the 0.9 m as part of the SMARTS (Small and Moderate
Aperture Research Telescope System) Consortium beginning in 2003
February.  The primary goals of CTIOPI are to discover and
characterize any nearby objects that remain unidentified in the solar
neighborhood --- primarily red dwarfs, brown dwarfs, and white dwarfs.
Although CTIOPI concentrates on southern hemisphere systems, we also
target important nearby star candidates as far north as $\delta =
+30$.  Most of the stellar systems observed during CTIOPI have been
selected because available astrometric data (high proper motions,
crude parallaxes) or photometric/spectroscopic distance estimates
indicate that they might be within 25 pc, the horizon of the Catalog
of Nearby Stars \citep{1991adc..rept.....G} and NStars
\citep{2003fst3.book..111H} compendia.  Objects possibly nearer than
10 pc are given highest priority.  The parallaxes we measure typically
have errors less than 3 mas, corresponding to distance errors of only
1--3\% for systems within 10 pc.  At the beginning of 2006, we
continued to observe roughly two-thirds of the more than 400 systems
targeted for parallax measurements during the 0.9 m program.  Under
SMARTS, we have expanded CTIOPI to carry out a program to search for
low mass companions to nearby stars, called ASPENS (Astrometric Search
for Planets Encircling Nearby Stars), in collaboration with David
Koerner at Northern Arizona University.

This paper is the fourth publication in {\it The Solar Neighborhood}
series that includes CTIOPI parallaxes. \citet{2005AJ....129.1954J}
presented 46 parallaxes for high proper motion systems observed at the
0.9 m.  \citet{2005AJ....130..337C} and Costa et al.~(2006, AJ in
press) presented 62 total parallaxes for fainter systems, including
brown dwarf candidates, observed at the 1.5 m.  In this paper, we
concentrate on systems observed at the 0.9 m that were candidates for
the 10 pc sample and present 27 total parallaxes.

\section{Photometry}
\label{sec:phot}

Because of the significant investment in observing time required to
determine a trigonometric parallax --- roughly one integrated night
over at least two years for each target --- it is prudent to obtain
characterization photometry of nearby star candidates before placing
them on the astrometric program.  As shown in
\citet{2004AJ....128.2460H}, the combination of $V_{J}$, $R_{KC}$ and
$I_{KC}$ (hereafter, without subscripts)\footnote{Subscript: J $=$
Johnson, KC $=$ Kron-Cousins. The central wavelengths for $V_{J}$,
$R_{KC}$ and $I_{KC}$ are 5475\AA, 6425\AA~and 8075\AA, respectively.}
photometry and infrared $J$, $H$, and $K_s$ photometry from 2MASS can
be used in a suite of color-absolute magnitude relations to estimate
distances of main sequence stars with 15\% accuracy.

$VRI$ photometry is reported in Table 2.  Two names are often given,
followed by the new optical $VRI$ photometry and the number of nights
on which observations were taken during CTIOPI.  Photometry was
acquired at the CTIO 0.9 m using the same filter and detector
combination used for the astrometry frames (see
section~\ref{sec:astr.obs} for details of the instrumental setup).
All observations were taken between November 1999 and December
2005.\footnote{The Tek \#2 $VRI$ filter set at CTIO has been used.
However, the Tek \#2 $V$ filter cracked in March 2005 and was replaced
by the very similar Tek \#1 $V$ filter.  Reductions indicate no
significant differences in either astrometric or photometric results
from the two filters.}  Data were reduced via IRAF with typical bias
subtraction and dome flat-fielding, using calibration frames taken at
the beginning of each night.  Standard star fields from
\citet{1982PASP...94..244G}, \citet{1990A&AS...83..357B}, and/or
\citet{1992AJ....104..372L} were observed several times each night to
derive transformation equations and extinction curves.  Apertures
14\arcsec~in diameter were used to determine the stellar fluxes to
match the apertures used by Landolt, except in cases when close
sources needed to be deconvolved (LP 771-095 ABC) or excised (LHS 288,
WT 460), in which case smaller apertures were used and aperture
corrections were done.  Further details about data reduction,
transformation equations, etc., can be found in
\citet{2005AJ....129.1954J}.

As discussed in \citet{2004AJ....128.2460H}, representative total
errors in the $VRI$ photometry are $\pm$ 0.03 mag.  Of the 81 $VRI$
photometric values reported here, 69\% have formal total 1$\sigma$
errors of 0.03 or less.  Only eight values have errors in excess of
0.05 mag: G 099-049, GJ 300, GJ 2130A, LHS 1610, and LHS 1723 at $V$
with errors of 0.05 to 0.07 mag, and WT 460 with errors of 0.09, 0.06
and 0.05 at $VRI$, respectively, because it is crowded by nearby
sources and required apertures of 5\arcsec~or smaller and large
aperture corrections.

A detailed variability study for the parallax targets has been carried
out by comparing the photometry of the target stars to the reference
stars that set the astrometric grid for each field.  Our methodology
matches that of \citet{1992PASP..104..435H}, where details of the
algorithm used can be found.  Columns 7--10 of Table 2 list the filter
used for parallax frames, the standard deviation of the target's
magnitude in that filter, the number of different nights the target
was observed, and the total number of frames evaluated for
variability.  In general, stars with standard deviation values
(relative to the reference stars) greater than 0.02 mag are obviously
variable, those with values 0.01--0.02 mag are variable at a level of
a few percent, and those with values less than 0.01 mag are ``steady.''
However, these cutoffs are not absolute because if the reference stars
used to measure the flux modulations are faint, an artificially high
standard deviation value may result.  Therefore, a `v'' is given after
the value in Column 8 if variability has been detected reliably.  In
total, we identify six of the 27 targets to be certain variables, a
rate of 22\%.  This is a lower than the 49\% rate (21 of 43 red
dwarfs) found to be variable by \citet{1994AJ....107.1135W}, primarily
because our variability threshold is roughly twice as high as that of
Weis.

Inspection of the photometric results indicates that many of the 20
new nearby systems are relatively bright, with eight systems having V
$<$ 13.  Only one of these systems, LP 771-095 ABC, was observed by
Hipparcos, but the close proximity of two sources, A and BC, resulted
in a trigonometric parallax with high error, 92.97 $\pm$ 38.04 mas.

Infrared photometry in the $JHK_{s}$ system (rounded to the hundredth)
has been extracted from 2MASS and is given in Columns 11--13 of Table
2.  Because of the stars' generally large proper motions, each
identification was confirmed by eye.  The $JHK_{s}$ magnitude errors
that include target, global, and systematic terms, i.e.~the {\it
x}$_{-}$sigcom errors (where {\it x} is j, h, or k), are almost always
less than 0.05 mag and are typically 0.02-0.03 mag.  The only two
exceptions are H band magnitude errors for LHS 337 (0.06 mag) and GJ
1207 (0.08 mag).

Column 14 of Table 2 provides a distance estimate and error for each
system, based on the suite of photometric distance relations in
\citet{2004AJ....128.2460H}. Column 15 lists the number of distance
relations valid and used for each target, where the maximum is 12.
These distances will be compared to the distances determined from the
trigonometric parallaxes reported in section~\ref{sec:astr.reconsnew}
to confirm or reveal new multiple systems, as discussed in
section~\ref{sec:disc.multiples}.

\section{Spectroscopy}
\label{sec:spec}

After the acquisition of optical and infrared photometry and the
subsequent distance estimates, a final spectroscopic check is
typically done to confirm the likely proximity of a candidate.
Spectral types are useful for the separation of dwarfs, subdwarfs, and
the occasional giant that slips through the photometric net.

As part of a long-term program to characterize nearby star candidates,
spectra were acquired between March 2002 and September 2004 on the
CTIO 1.5 m and 4.0 m telescopes.  On the 1.5 m telescope, observations
were made using a 2\farcs0~slit in the RC Spectrograph with grating
\#32 and order blocking filter OG570 to provide wavelength coverage
from 6000 to 9500 \AA~and resolution of 8.6 \AA~on the Loral 1200x800
CCD.  On the 4.0 m telescope, observations were made using a
2\farcs0~slit in the RC Spectrograph with grating G181 and order
blocking filter OG515 to provide wavelength coverage from 5000 to
10700 \AA~and resolution of 5.6 \AA~on the Loral 3K x 1K CCD.  At
least two sequential exposures were taken of each target to permit
cosmic ray removal.  Reductions were accomplished using standard
methods in IRAF, and mild fringing, a result of the rear-illuminated
CCD on the 4.0 m was removed with a tailored IDL routine.

As shown in the last column of Table 2, all of the new nearby stars
are red dwarfs, with types spanning M3.0V to M8.5V.  What is somewhat
surprising is that many of the new systems are {\it not} late-type M
dwarfs --- 16 of the 20 primaries in the new RECONS systems have
spectral types earlier than M6.0V.  Thus, there apparently remain many
mid-type M dwarfs near the Sun that are not yet recognized as solar
neighbors.

\section{Astrometry Observations and Reductions}
\label{sec:astr.obs}

The 0.9 m telescope is equipped with a 2048 $\times$ 2048 Tektronix
CCD camera with 0\farcs401 pixel$^{-1}$ plate scale that has remained
on the telescope throughout CTIOPI.  All observations of the parallax
targets (``pi stars'') listed in Table 3 (coordinates in epoch and
equinox 2000.0) were made through $VRI$ filters using the central
quarter of the chip, yielding a 6\farcm8 square field of view.  The
same filter (Column 7 of Table 2 and Column 4 of Table 3) is used for
all astrometric observations for a given pi star.  The basic data
reduction for the astrometry CCD frames includes overscan correction,
bias subtraction and flat-fielding, using calibration frames taken at
the beginning or end of each night.

An extensive discussion of the data acquisition and reduction
techniques is given in \citet{2005AJ....129.1954J}.  Briefly, 5--10
parallax frames are typically taken for a parallax field at each epoch
to provide multiple measurements of pi and reference star centroids.
Frames are usually taken within $\pm$30 minutes of a pi star's transit
in order to minimize the corrections required for differential color
refraction (DCR), with the goal of having some frames on both sides of
the meridian on each night that a field is observed.  All observations
for a given field are made with a set of reference stars suitably
positioned on the chip within a few pixels of their original
locations.  Good reference star configurations include 5--15 stars
surrounding the pi star (the number of reference stars is given in
Column 9 of Table 3).  Typically, selected reference stars have at
least 1000 counts at the peak and minimal proper motion and parallax
(the ensemble of proper motion and parallax values for reference stars
is set to zero during reduction).  These reference stars are used as
the grid against which the pi star astrometry is measured, to correct
for any field translation or rotation caused by temporal shifts in the
telescope/camera combination, and as comparison stars for variability
studies.  In addition, $VRI$ photometry is obtained for the reference
stars and used to make DCR corrections and to estimate distances to
each star photometrically to permit the conversion from relative to
absolute parallax.

To decouple parallactic and proper motions in the final astrometric
solution, observations are typically taken over at least two years,
including four high parallax factor seasons.  Columns 5--8 of Table 3
list the number of seasons, total number of frames, epochs of
observation, and length of time for each pi star frame series.  As in
previous papers, in the seasons column, ``c'' indicates a continuous
set of observations throughout the seasons and ``s'' indicates
scattered observations, meaning that the target was observed on only
one night during one or more seasons (or was missed entirely during a
season); continuous observations are superior to scattered
observations.  The set of observations for a given pi star usually
includes at least 20 evening and 20 morning frames.  However, because
the pi stars in this paper were likely to be within 10 pc, they were
given high priority status throughout their tenures on the observing
list; thus, the total number of frames given for each target often far
exceeds the typical 40 total frames for a pi star.

\section{Astrometry Results}

The trigonometric parallax results are given in Table 3, including
both relative (Column 10) and absolute (Column 12) parallaxes, as well
as the the correction between the two (Column 11).  The corrections
are generally less than $\sim$2 mas, so systematics in the corrections
should not significantly affect the final results.  Proper motions and
position angles of the proper motions are given in Columns 13 and 14,
and the derived tangential velocity is given in Column 15.  The final
column indicates whether previous trigonometric values were available,
with notes following the table.  After outlining the definition of a
RECONS system, we discuss each system individually.

\subsection{Definition of RECONS Systems}
\label{sec:astr.reconsdefn}

Stars targeted in this study include potential new 10 pc sample
members with no trigonometric parallaxes and stars supposedly within
10 pc with large trigonometric parallax errors.  The second category
of targets allows us the opportunity to evaluate the threshold of
reliability for available trigonometric parallaxes.  As supported by
the results described below, we formally define the parameters for
inclusion in the RECONS 10 pc sample as: {\it A system comprised of
stars, brown dwarfs, and/or planets (and any associated surrounding
material) for which a trigonometric parallax of at least 100 mas has
been determined, and for which the parallax has a formal error of 10
mas or less.}

The advantages of this definition are (1) it assigns a clear ``in'' or
``out'' status for each system, rather than using probabilities for
systems, in particular those near the 10 pc border; (2) it establishes
a uniform 10 mas error cutoff for every system in the sample; and (3)
it provides a limit on the ``worst case'' for any system because the
distance is known to better than 10\% in all cases.  This final
benchmark is superior to current photometric or spectroscopic distance
estimates that, when done correctly, are not better than 15\%
\citep{2004AJ....128.2460H} for red dwarfs, the dominant component of
the nearby star population.

In truth, we do not consider a trigonometric parallax ``definitive''
unless its error is less than 3 mas, and this value may change when
large numbers of stars are measured more accurately via future
efforts.  However, we include systems with parallax errors of 3--10 in
the RECONS sample to bolster the statistics until we can acquire
improved data, as is our goal during CTIOPI.  We {\it do not} include
systems with parallax errors in excess of 10 mas, because, as shown in
section~\ref{sec:astr.reconsnot}, many systems with errors this large
do not prove to be within 10 pc.

One awkward aspect of the adopted definition is that it excludes
systems with very large parallaxes and large parallax errors until
accurate data are in hand.  Examples include stars with approximate
parallaxes measured using a few photographic plates, such as SO
0253+1652 and SCR 1845-6357 AB, and systems with $\pi_{trig}$ from
more extensive plate series having errors slightly in excess of 10
mas, such as G 041-014 ABC and LHS 288, both of which were reported to
be within five parsecs (see section~\ref{sec:astr.reconsnew}).  LHS
288 is a particularly subtle example of applying the RECONS criteria
for membership.  \citet{1986PASP...98..658I} determined a parallax of
221 $\pm$ 8 mas, but the YPC, which carries out a systematic
assessment of parallaxes and errors, gives an error of 11.3 mas for
LHS 288's parallax.  Thus, this star did not meet the formal
definition for a RECONS member because we used the YPC for sample
definition.  These ``nearly certain'' members were generally observed
early and intensely in CTIOPI to bring them formally into the sample.

\subsection{New RECONS Systems}
\label{sec:astr.reconsnew}

Here we report 20 new RECONS stellar systems based on the
trigonometric parallaxes given in the top portion of Table 3.
Previously, none of these systems had trigonometric parallaxes with
errors less than 10 mas, and 12 had no trigonometric parallaxes of any
kind.  Of the 20 new systems, five have proper motions in excess of
1\farcs0/year, 11 have motions between 1\farcs0/year and
0\farcs5/year, and four have motions less than 0\farcs5/year.  These
numbers indicate that there are likely to be large numbers of stars
within 10 pc of the Sun that have low, but not negligible, proper
motions.

Many of these systems have been recovered through our extensive
efforts to estimate reliable distances for stars included in a large
collection of potentially nearby objects.  The collection includes
various types of astrometric, photometric, and spectroscopic data that
have been converted to standard systems, evaluated for reliability,
and then used to pinpoint systems potentially nearer than 10 pc.
Primary sources include (1) the valuable Catalog of Nearby Stars
\citep[ hereafter CNS]{1991adc..rept.....G}, which includes many stars
with photometric parallaxes, (2) the Luyten Half Second Catalogue
\citep{1979lccs.book.....L}, which includes 3602 objects with proper
motions in excess of 0$\farcs$5 yr$^{-1}$, and which still accounts
for 87\% of all known such objects as of this writing
\citep[see][]{2005AJ....130.1658S}, and (3) the monumental work of
Weis and collaborators, who provided accurate optical photometry and
computed photometric parallaxes for thousands of stars
\citep{1984ApJS...55..289W, 1986AJ.....91..626W, 1987AJ.....93..451W,
1988AJ.....96.1710W, 1988PASP..100..749B, 1991AJ....101.1882W,
1991AJ....102.1795W, 1993AJ....105.1962W, 1996AJ....112.2300W}.

More recently, large scale work by Reid and collaborators
\citep{2002AJ....123.2806R, 2002AJ....123.2822R, 2002AJ....123.2828C,
2003AJ....125..354R, 2003AJ....126.2421C, 2003AJ....126.2449R,
2003AJ....126.3007R, 2004AJ....128..463R} has identified additional
new systems from proper motion compendia, and has reidentified and
confirmed many of the systems noted by others.  In addition, our own
new recent proper motion survey, known as the SuperCOSMOS-RECONS (SCR)
effort \citep{2004AJ....128..437H, 2004AJ....128.2460H,
2005AJ....129..413S, 2005AJ....130.1658S} has already provided three
new systems described in this paper.  The initial dates for the
astrometric series listed in Table 3 indicate when the targets were
first observed, illustrating that many targets were placed on the
parallax program before most of the more recent distance estimating
efforts.

Here we provide a brief dossier for each new RECONS sample system.
The notes to Table 3 list previously available trigonometric
parallaxes, which are also given here in the text along side
photometric or spectroscopic distances estimates.  It is also
important to note that the trigonometric parallaxes given in the 1991
edition of the CNS are from a preliminary version of the YPC.  Thus,
they are of the same origin as the definitive YPC, published in 1995,
and are therefore not listed explicitly here.  The final YPC
parallaxes often have somewhat larger errors determined systematically
during construction of the catalog, and errors in excess of 10 mas
exclude some stars from the RECONS sample until the present study,
even though their CNS errors were less than 10 mas.

{\bf LHS 1302} was reported to have photometric distances of 10.0
$\pm$ 1.8 pc in CNS and 10.2 $\pm$ 1.3 by \citet{2002AJ....123.2806R}.
The distance is 9.92 $\pm$ 0.19 pc, consistent with previous estimates
and near the horizon of the RECONS sample.  There is a hint of a
perturbation in the RA residuals that may be confirmed or refuted
during via ASPENS.  The star is variable at a level of 0.05 mag in the
$R$ band during 26 nights of observation, and flared by 0.13 mag in R
on UT 2004 November 22 during a sequence of ten frames spanning 37
minutes.

{\bf APMPM J0237$-$5928} is an X-ray source reported by
\citet{1999A&A...345L..55S} to have a photometric distance between 11
and 14.5 pc based on a spectral type of M5V.  The distance is 9.64
$\pm$ 0.10 pc.  The star is mildly variable, with a standard deviation
of 0.013 mag in the $R$ band during 27 nights of observation, with a
full variability range of 0.06 mag.

{\bf SO 0253$+$1652} was reported by \citet{2003ApJ...589L..51T}, to
have a crude trigonometric parallax of 410 $\pm$ 90 mas (2.4 $\pm$ 0.6
pc) that was described as a ``lower limit on the true parallax.''
Based on this value, claims were made that it was the third nearest
star system, was an extremely metal poor subdwarf, or had a radius
only 60\% that of a comparable star, GJ 1111.  However, their
photometric estimate (3.6 $\pm$ 0.4 pc) and our group's estimate
\citep[3.7 $\pm$ 0.6 pc,][]{2004AJ....128.2460H} are better matches to
the actual distance of 3.84 $\pm$ 0.04 pc determined here.  Rather
than being an exotic object, SO 0253+1652 is a normal red dwarf with
$M_V = 17.22$ and spectral type M7.0V, and is the 23rd nearest system
to the Sun.

{\bf LP 771$-$095 ABC} is a triple system (A is LP 771-095 and the
close pair BC is LP 771-096) with separations of 7\farcs22~at PA $=$
315$^\circ$ for A--BC \citep{2003AJ....125..332J} and 1\farcs30~at PA
$=$ 138$^\circ$ for B--C (this paper), with no change in position
angle noted during the six years of our observations.  A and BC have
separate photometry, with A brighter than BC at $V$ and $R$, but
fainter at $I$ (see Table 2).  The magnitude differences between B and
C are $\Delta$V $= 0.86 \pm 0.09$, $\Delta$R $= 0.75 \pm 0.03$, and
$\Delta$I $=0.66 \pm 0.07$.

The CTIOPI reference field is faint and sparse, and as few as four
reference stars were used to boost the frame count.  Nevertheless, we
measure consistent parallaxes for A and BC, although the parallax
error for BC (4.99 mas) is large because of elongated images due to
its duplicity that result in poorly determined centroids.  The
weighted mean of the two parallaxes is 144.68 $\pm$ 2.52 mas (6.91
$\pm$ 0.12 pc), a factor of 15 improvement in the 92.97 $\pm$ 38.04
mas (10.8 $\pm$ 5.3 pc) value measured by Hipparcos.  The system was
reported to have photometric distances of 7.6 $\pm$ 1.8 pc (for A and
BC) in CNS, 7.8 pc in \citet{1991AJ....102.1795W}, and 10.3 pc for A
and 12.1 pc for BC by \citet{2004AJ....128..463R}.  After deconvolving
BC, the latter authors estimate a distance of 10.2 pc.  The system is
actually closer than all photometric estimates.

On UT 1999 August 22, the BC pair showed a 0.30 mag decrease in
brightness.  No other night shows such a decrease, so we cannot
confirm that this is an eclipsing event, but note it for future
monitoring.  Neither the A component, nor any of the reference stars
show this anomaly.  The standard deviations in variability for both A
and BC are artificially high because the reference stars are faint ---
neither source is obviously variable.

{\bf LHS 1610} has a trigonometric parallax in YPC of 70.0 $\pm$ 13.8
mas (14.3 $\pm$ 2.9 pc).  The parallax value and its large error both
exclude the star from the RECONS sample.  LHS 1610 was reported to
have photometric distances of 9.6 $\pm$ 1.4 pc in CNS and 10.5 $\pm$
1.2 pc by \citet{2002AJ....123.2806R}.  The distance is 9.85 $\pm$
0.20 pc, consistent with the photometric estimates and near the
horizon of the RECONS sample.

The star is obviously variable, with a standard deviation of 0.024 mag
in the $V$ band during 23 nights of observation.  Parceling the data
into individual nights shows that the star varies by 0.08 mag,
possibly regularly.  Continuing observations as part of ASPENS will
provide additional epochs of photometry to further investigate the
periodicity.

{\bf LHS 1723} was reported to have photometric distances of 6.1 $\pm$
1.0 pc in CNS, 9.2 pc by \citet{1994AJ....108.1437H}, 9 pc by
\citet{1998AJ....115.1648P}, and 5.7 $\pm$ 0.5 pc by
\citet{2002AJ....123.2822R}.  The distance is 5.32 $\pm$ 0.04 pc.  The
star is obviously variable, with a standard deviation of 0.020 mag in
the $V$ band during 27 nights of observation, with a full variability
range of 0.09 mag.

{\bf G 099$-$049} has a trigonometric parallax in YPC of 186.2 $\pm$
10.1 mas (5.4 $\pm$ 0.3 pc).  The large error excludes the star from
the RECONS sample until the present study.

The formal correction from relative to absolute parallax is 4.50 $\pm$
0.82 mas because the reference stars are red.  The field is in Orion,
so the redness is artificial, thereby causing the reference stars to
appear much nearer than they are, with photometric parallaxes of 3--17
mas.  The reference stars' trigonometric parallaxes and proper motions
are well-behaved, being less than 1 mas and 10 mas/yr, respectively.
We have therefore adopted a reasonable ``mean" correction to absolute
parallax of 1.50 $\pm$ 0.50 mas (assuming a large error) instead of
the erroneous formal correction.

The star exhibits a possible perturbation of the photocenter at a
level of $\sim$20 mas in the RA residuals; further investigation
during ASPENS will help confirm or refute the perturbation.  The
source 6\arcsec~to the NW in late 2005 is a background source, and may
be affecting the centroids.  Our derived proper motion, 0\farcs313/yr,
and direction of the proper motion, 98$^\circ$, are significantly
different than given in CNS, 0\farcs241/yr at 108$^\circ$.
\citet{2005AJ....129.1483L} report a proper motion of 0\farcs314/yr at
98$^\circ$, consistent with our value.

{\bf SCR 0630$-$7643 AB} was first reported in
\citet{2004AJ....128.2460H} to be a potential nearby binary of type
M6.0VJ with an estimated photometric distance of 7.0 $\pm$ 1.2 pc,
assuming magnitude differences in $VRIJHK$ of 0.25 mag.
\citet{2005AJ....129..413S} estimated a distance of 6.9 pc based on
photographic plate and 2MASS magnitudes.

The separation of AB is 0\farcs90~at PA $=$ 34$^\circ$, with no change
in position angle noted during the two year period of the
observations.  The magnitude differences between A and B are $\Delta$V
$= 0.20 \pm 0.03$, $\Delta$R $= 0.23 \pm 0.06$, and $\Delta$I $=0.21
\pm 0.10$, thereby confirming that the two stars are similar in
brightness and color.  Frames with seeing better than 1\farcs2~were
removed from the reduction because the images are elongated.  The
trigonometric parallax reported here corresponds to a distance of 8.76
$\pm$ 0.14 pc, further than previous estimates.  The projected
separation is 7.9 AU, which implies an orbital period of nearly 50 yr
for two stars with masses of 0.10 M$_\odot$ each.

{\bf G 089$-$032 AB} was reported to be a single star with photometric
distances of 5.8 pc by \citet{1986AJ.....91..626W}, 6.2 $\pm$ 1.0 pc
in CNS, and 8.1 pc by \citet{1994AJ....108.1437H}.  The distance is
8.58 $\pm$ 0.07 pc --- the distance estimates were generally too low
because the system is a close double.

The system was first reported to be a binary with separation
0\farcs7~and of nearly equal brightness at $K$ by
\citet{1999ApJ...512..864H}.  During CTIOPI, the resulting FWHM
measurements are typically $>$ 1\farcs5~because the unresolved B
component stretches the images.  The components are too close together
for any clear measurements of magnitude differences in CTIOPI frames,
but we can estimate the position angle of the companion to be
$\sim$300$^\circ$, with no obvious change in the six years of
coverage.  At the measured distance, the projected separation is 6.0
AU, and implies an orbital period of $\sim$30 years, assuming masses
of 0.13 M$_\odot$ for each component.  No perturbation can yet be
identified in the RA and DEC residuals, consistent with the long
orbital period.

{\bf GJ 300} has a trigonometric parallax in YPC of 169.9 $\pm$ 15.0
mas (5.9 $\pm$ 0.5 pc).  The large error excludes the star from the
RECONS sample until the present study.  \citet{2004AJ....128..463R}
reported a photometric distance of 6.3 pc.  We measure a distance of
7.96 $\pm$ 0.06 pc, which is inconsistent with the previous
trigonometric measurement and indicates that the uncertainty was
optimistic.  The star is mildly variable, with a standard deviation of
0.016 mag in the $V$ band during 31 nights of observation, with a full
variability range of 0.06 mag.

{\bf G 041$-$014 ABC} is a triple system in which A and B form a close
spectroscopic binary, and C is more distant.
\citet{1997AJ....113.2246R} report a separation of 1 AU for AB, while
\citet{1999A&A...344..897D} report an orbital period of 7.6 days.  For
component masses of 0.19 and 0.17 M$_\odot$ for A and B (estimated
after deconvolving the C component, then using the ratio of the
spectroscopic $K_1$ and $K_2$ values and assigning masses via the
mass-luminosity relation of \citet{1999ApJ...512..864H}), the
separation would be only 0.05 AU.  Nonetheless, the system is
photometrically steady, with a standard deviation of only 0.009 mag in
the $V$ band during 22 nights of observation, indicating no obvious
interaction between the components.

\citet{1999A&A...344..897D} imaged component C at a separation of
0\farcs62~at PA $\sim$ 90$^\circ$ in 1997.  The stars cannot be
resolved in CTIOPI frames, so the photometry is for the combined light
of all three stars.  In fact, no elongation is seen even in frames
with images having FWHM of 1\farcs0.  At the measured distance, the
projected separation of AB--C is 4.2 AU, implying an orbital period of
11.5 years, assuming an additional mass of 0.17 M$_\odot$ for
component C.  A hint of a perturbation at a level of 10 mas over six
years can be seen in the RA residuals, but much more data are needed
to confirm it.  This perturbation is consistent with an orbital period
that may be much longer than that derived from the projected
separation.

The system was reported to have photometric distances of 4.5 $\pm$ 0.7
pc in CNS, 4.6 pc by \citet{1991AJ....102.1795W}, and 5.3 pc by
\citet{1994AJ....108.1437H}, all of whom assumed it to be a single
star.  The distance is 6.77 $\pm$ 0.09 pc --- the distance estimates
were too low because the system is a close triple.

{\bf LHS 2090} was reported to have photometric distances of 6.0 $\pm$
1.1 pc by \citet{2001A&A...374L..12S}, 5.2 $\pm$ 1.0 by
\citet{2002AJ....123.2806R}, and 5.7 $\pm$ 0.9 pc by
\citet{2004AJ....128.2460H}.  The distance is 6.37 $\pm$ 0.11 pc,
consistent with previous estimates.

{\bf LHS 2206} was reported to have a photometric distance of 10.2
$\pm$ 1.6 pc in CNS.  The distance is 9.23 $\pm$ 0.20 pc, consistent
with the previous estimate, and firmly placing the star in the RECONS
sample.

{\bf LHS 288} is a well-known nearby star in a very crowded field with
a trigonometric parallax in YPC of 222.5 $\pm$ 11.3 mas (4.5 $\pm$ 0.2
pc).  The large error excludes the star from the RECONS sample until
the present study.  However, the original result by
\citet{1986PASP...98..658I}, 221 $\pm$ 8 mas (4.5 $\pm$ 0.2 pc), has
an error small enough to include the star in the RECONS sample.
Fortunately, Ianna is an author on this paper as well as the original
parallax paper, so can be credited with the discovery of LHS 288 as a
RECONS system, whether the date is 1986 or at time of this paper.
\citet{2004AJ....128.2460H} reported the star to have a photometric
distance of 6.9 $\pm$ 1.7 pc.  We measure a distance of 4.79 $\pm$
0.06 pc, consistent with previous measurements.

{\bf SCR 1138$-$7721} was first reported in
\citet{2004AJ....128..437H} to be a potential nearby star with an
estimated photometric distance based on plate and 2MASS magnitudes of
8.8 $\pm$ 2.7 pc (when both internal and external errors are
included).  \citet{2004AJ....128.2460H} provided a distance estimate
of 9.4 $\pm$ 1.7 pc using CCD and 2MASS magnitudes.  The distance is
8.18 $\pm$ 0.20 pc, consistent with the previous estimates.

{\bf LHS 337} was reported to have photometric distances of 7.2 $\pm$
1.5 pc in CNS and 9 pc by \citet{1998AJ....115.1648P}.  The distance
is 6.38 $\pm$ 0.08 pc.

{\bf WT 460} was reported to have a photometric distance of 11 pc by
\citet{1998AJ....115.1648P}.  The distance is 9.31 $\pm$ 0.13 pc,
firmly placing the star in the RECONS sample.

The star is corrupted by two sources within 6\arcsec~in the NE
quadrant, neither of which is a common proper motion companion.
Astrometric residuals with RMS values of 24 mas in RA and 32 mas in
DEC are roughly three times typical residuals, but the large number of
frames and strong reference star configuration allow an accurate
parallax to be determined.  This star will not be followed in ASPENS
because of the corrupting background sources.

{\bf GJ 1207} has a trigonometric parallax in YPC of 104.4 $\pm$ 13.6
mas (9.6 $\pm$ 1.3 pc).  The large error excludes the star from the
RECONS sample until the present study.  We measure a distance of 8.67
$\pm$ 0.11 pc, which is consistent with the previous measurement.

GJ 1207 has the largest photometric standard deviation --- 0.263 mag
in the $V$ band during 27 nights of observation --- of any star
reported here, primarily because of one extreme flare event.  On UT
2002 June 17, the star's brightness was normal relative to reference
stars in the first frame, then increased by a factor of 4.6 (1.7 mag)
by the second frame, and subsequently faded to a factor of 2.5 (1.0
mag) higher than normal by the sixth frame.  The entire sequence
spanned only 15 minutes.

{\bf SCR 1845$-$6357 AB} was first reported in
\citet{2004AJ....128..437H} to be a very red, potential nearby star
with an estimated photometric distance based on plate and 2MASS
magnitudes of 3.5 $\pm$ 1.2 pc (when both internal and external errors
are included).  A spectral type of M8.5V and distance estimate of 4.6
$\pm$ 0.8 pc were presented using CCD and 2MASS magnitudes in
\citet{2004AJ....128.2460H}.  \citet{2005AJ....129..409D} measured a
preliminary trigonometric parallax, 282 $\pm$ 23 mas (3.5 $\pm$ 0.3
pc), based on eight SuperCOSMOS photographic plates, illustrating that
reasonable parallaxes can be derived with only a few archival plates.
We measure a parallax of 259.45 $\pm$ 1.11 mas (3.85 $\pm$ 0.02 pc),
consistent with all previous estimates.  This distance makes SCR
1845-6357 AB the 24th nearest system to the Sun.

\citet{2006ApJ...641L.141B} report a T dwarf companion separated by
1\farcs17~at position angle 170$^\circ$ in 2005.  We find a very low
error in the parallax because of intense coverage and a rich,
well-balanced set of reference stars.  We do not yet see any curvature
in the the astrometric centroids due to the companion, but the
residuals do show a larger scatter than for most stars, possibly due
to the pull of the companion.  Assuming masses of 0.10 and 0.05
M$_\odot$ and a 4 AU separation, we derive a 21-year orbital period.
The photocentric semimajor axis is predicted to be $\sim$350 mas,
corresponding to a perturbation in the photocenter of $\sim$34
mas/year, although the rate depends on the true size, shape, and
orientation of the orbit on the sky.  Continued observations are
planned to improve and monitor the position of the center of mass of
the system.

{\bf LHS 3746} was reported to have photometric distances of 10.0
$\pm$ 2.7 pc in CNS and 10 pc by \citet{1998AJ....115.1648P}.  The
distance is 7.45 $\pm$ 0.07 pc.  The relative photometry shows a
long-term trend possibly indicative of a stellar activity cycle.

\subsection{Known RECONS Systems}
\label{sec:astr.reconsknown}

{\bf GJ 54 AB} is a fast-orbiting pair of red dwarfs that
\citet{1974PASP...86..742R} reported as a ``Suspected very close
binary in the 40-inch reflector.  The components are assumed to be of
equal magnitude.'' No separation, position angle, or epoch were given,
and the separation must have been near the diffraction limit of the
Siding Spring Observatory 40-inch.  \citet{2004AJ....128.1733G}
provide details of the system, including clear resolution of the pair
using HST-NICMOS.  The system has been followed since 2000 in our
continuing HST-FGS effort to measure accurate masses for low mass
stars \citep{2004ASPC..318..159H}.  The $HST-FGS$ measurements
indicate that $\Delta$$V =$ 1.04, and a preliminary orbital analysis
yields an orbital period of $\sim$1.1 years and semimajor axis of 0.9
AU.

As can be seen in Figure~\ref{fig:henry1}, the perturbation affects
the astrometric residuals in both RA and DEC (9.1 and 13.6 mas RMS,
respectively) after solving for proper motion and parallax.  For
comparison, the ``well-behaved'' GJ 300 RA and DEC residuals (3.6 mas
RMS in each axis) are shown.  For GJ 54 AB, beating of the 1.0 year
parallax motion and 1.1 year orbital motion makes deconvolution of the
two motions problematic, and a high parallax error (3.38 mas) results
(and is not helped by a faint set of reference stars, which also cause
the variability measurement to be artificially high).  In fact, this
is nearly a worst-case scenario because of the confluence of the two
perturbations on the star's straight-line proper motion path, coupled
with the small semimajor axis of the photocenter's orbital motion,
only 16 mas.  The system has trigonometric parallax measurements
listed in YPC and HIP of 120.5 $\pm$ 10.1 mas (8.30 $\pm$ 0.70 pc) and
122.86 $\pm$ 7.53 mas (8.14 $\pm$ 0.50 pc), respectively.  Our data
yield a better parallax, but the weighted mean of the three
independent parallax measurements, 136.62 $\pm$ 2.95 mas (7.32 $\pm$
0.16 pc), still has a relatively high error.  The system will be
observed in ASPENS to improve the parallax, monitor the motion of the
photocenter, and facilitate the derivation of accurate masses.

{\bf GJ 1061} was reported to have a photometric distance of 4.3 $\pm$
1.1 pc in CNS.  \citet{1997AJ....114..388H} discovered the star to be
the 20th nearest stellar system with a trigonometric parallax of 273.4
$\pm$ 5.2 mas (3.66 $\pm$ 0.07 pc) from photographic plates.  Three
other stars of similar nearby pedigree reported in that work turned
out to be distant giants, providing classic illustrations of how
photometry and spectroscopy can be used to fine-tune astrometry
observing lists, before spending observing time on stars that will
have negligible parallax.  Here we provide a CCD parallax of 271.92
$\pm$ 1.34 (3.68 $\pm$ 0.02 pc) for GJ 1061, which has an error four
times smaller than the previous value and is completely consistent.
GJ 1061's remains the 20th nearest system.

\subsection{Not RECONS Systems}
\label{sec:astr.reconsnot}

{\bf GJ 1001 ABC} is a triple system with separations of 18\farcs2~at
PA $=$ 259$^\circ$ for A--BC and a changing separation always less
than 0\farcs1~for B-C \citep{2004AJ....128.1733G}.  The initial L
dwarf companion was reported by the \citet{1999A&A...351L...5E}, and
was resolved by \citet{2004AJ....128.1733G} using HST-NICMOS into two
nearly equal magnitude L dwarfs.  The BC pair is evident in the
parallax frames as a very faint source, but no $\pi_{trig}$
measurement was attempted.

GJ 1001 A has a trigonometric parallax in YPC of 104.7 $\pm$ 11.4 mas
(9.6 $\pm$ 1.1 pc).  The distance determined here is 13.01 $\pm$ 0.67
pc, clearly pushing the system beyond the horizon of the RECONS
sample.  So far, the improvement in the parallax error is only a
factor of three because of a set of faint reference stars and poor
observational coverage.  The primary will remain a CTIOPI target
because an accurate parallax for the system offers the opportunity to
determine crucial masses for the brown dwarfs, B and C, which
continuing observations show have an orbital period of $\sim$4 years
(Golimowski et al.~in prep).

{\bf GJ 633} has a trigonometric parallax in YPC of 104.0 $\pm$ 13.7
mas (9.6 $\pm$ 1.3 pc).  The distance determined here is 22.48 $\pm$
0.93 pc, well beyond the 10 pc limit of the RECONS sample.  A
photocenter perturbation of $\sim$60 mas is seen in the RA residuals,
which may be real or could be due to a background star 7\arcsec~to the
E that corrupts centroid measurements.  The distance derived using
$VRIJHK$ photometry and the suite of relations in
\citet{2004AJ....128.2460H} is 19.5 $\pm$ 3.0 pc, which is consistent
with the trigonometric distance.  The target will remain on the
program to determine the veracity of the perturbation and the
existence of a possible companion that would contribute minimal light
to the system.

{\bf GJ 2130 ABC} is a triple system with a separation of
21\farcs23~at PA $=$ 88$^\circ$ for A--BC \citep[ incorrectly noted as
AC and B]{2003AJ....125..332J}.  BC is reported to be a double-lined
spectroscopic binary by \citet{1999A&A...344..897D}.  The duplicity of
BC is not evident in frames with images having FWHM of 1\farcs0.

\citet{1979A&AS...38..423G} provided distance estimates of 14.5 pc for
A and 19.1 pc for B (C was not known at the time).
\citet{2004AJ....128..463R} provided distances of 13.5 pc for A and
6.2 pc for BC (adopted from HIP).  The distance estimates for the
presumed single star, A, are good matches to our consistent
measurements of parallaxes for A and BC, corresponding to 14.00 $\pm$
0.52 pc and 14.24 $\pm$ 0.42 pc, respectively.  The weighted mean of
the two parallaxes for the system is 70.69 $\pm$ 1.62 mas (14.15 $\pm$
0.32 pc), which is a factor of 14 improvement in the value measured by
Hipparcos, 161.77 $\pm$ 11.29 mas (6.2 $\pm$ 0.4 pc), and a factor of
five improvement in the value of 68.5 $\pm$ 7.9 mas (14.6 $\pm$ 1.7
pc) found by \citet{2000A&AS..144...45F}, who used Hipparcos transit
data to improve the result.

\section{Discussion}
\label{sec:discussion}

\subsection{The Solar Neighborhood Population}
\label{sec:disc.population}

After the publication of the YPC in 1995 and Hipparcos results in
1997, there were 215 systems in the RECONS sample.  Between the time
of those publications and the first parallax results from CTIOPI
(2005), only eight new systems had reliable trigonometric parallaxes
measured that placed them within 10 pc (see Table 1), and the nearest,
GJ 1061, was by our group \citep{1997AJ....114..388H}.  CTIOPI has
already reported eight additional new 10 pc systems meeting the RECONS
criteria --- five in \citet{2005AJ....129.1954J}, two in
\citet{2005AJ....130..337C}, and one in Costa et al.~(2006, AJ in
press).  In this paper, we add an additional 20 systems to the sample
of stars within 10 pc with high quality parallaxes.  Counting
additions in this paper, since the publication of YPC and HIP results,
the total number of new RECONS systems stands at 36.  The 34 systems
added since 2000 constitute a 16\% increase in the number of systems
within 10 pc in six years.

Figure~\ref{fig:henry2} shows an HR diagram of the RECONS sample,
plotting $M_V$ vs. $V-K$ for all systems with parallaxes meeting the
RECONS criteria (see Section~\ref{sec:astr.reconsdefn}).  All
multiples have been deconvolved into individual components as well as
can currently be done at both the $V$ and $K$ bandpasses.  Small
circles indicate objects known to be within 10 pc after publication of
the YPC and HIP catalogs, as well as new discoveries by other groups.
Solid points represent RECONS discoveries.  

Labels are given for prominent or unusual stars.  Triangles mark the
four known subdwarfs within 10 pc.  AU Mic and AT Mic AB (GJ 803 and
GJ 799 AB, respectively) form a wide triple system that is apparently
quite young.  The system appears to be comoving with the $\beta$ Pic
group and has an age of only $\sim$20 Myr.  The position of all three
stars above the main sequence supports the hypothesis that the system
is young, and the stars have not yet descended onto the main sequence.
Further support came recently, when AU Mic was found to be the nearest
(9.9 pc) star with a dust disk directly observed at optical and
near-infrared wavelengths \citep{2004Sci...303.1990K}.  The three
known L dwarfs within 10 pc --- 2MA 0036+1821, 2MA 1507-1627, and DEN
0255-4700 --- are shown, but none of the eight known T dwarfs or four
extrasolar planets are plotted.

All of the systems reported in this paper are mid- to late-type M
dwarfs (spectral types listed in Table 2).  The three brightest new
stars are labeled --- LP 771-095 A, LHS 3746, and GJ 1207 --- as well
as many of the fainter discoveries from the RECONS group and others.
Note in particular the $M_V$ of 24.4 for DEN 0255-4700, for which a
parallax and photometry were determined during our CTIOPI 1.5m program
(Costa et al.~2006, AJ in press).  We believe that this is the
faintest object outside the Solar System for which $M_V$ has been
determined.

The dominance of the nearby stellar population by red dwarfs is
obvious from Figure~\ref{fig:henry2}.  For objects in the refereed
literature, including this paper, the complete count of current
objects meeting the RECONS criteria is: 18 white dwarfs, 4 A types, 6
F, 21 G, 44 K, 240 M, 3 L, 8 T, and 4 extrasolar planet candidates.
Counts include the Sun and the adoption of a dividing line between
spectral types K and M at $M_V$ $=$ 9.00.  Thus, of the 348 total
objects, the 240 M dwarfs comprise 69\% of the nearby population.
Elimination of the presumably non-stellar L, T, and extrasolar planet
portions brings the fraction to 72\% of all stars, and this fraction
will continue to grow as more nearby red dwarfs are revealed.

\subsection{Hidden Multiples}
\label{sec:disc.multiples}

To confirm or reveal unresolved multiple systems, we make a comparison
between the accurate trigonometric distances presented here and in
previous CTIOPI papers and the photometric distances estimated via the
suite of $M_K$-color relations described in
\citet{2004AJ....128.2460H}.  This technique provides distance
estimates good to 15\%, measured by running RECONS stars with accurate
trigonometric parallaxes back through the suite of relations, under
the assumption that the stars are single.  Photometric distance
estimates for the systems discussed in detail in this paper are given
in column 14 of Table 2, and the number of relations used for each
target (maximum of 12) is given in column 15.

Figure~\ref{fig:henry3} compares the photometric distances with
trigonometric distances for objects within 20 pc, as reported by
CTIOPI in the four papers with parallaxes to date.  The distance
cutoff of 20 pc has been selected because that is the distance limit
of our 20-20-20 sample (within 20 pc, orbital periods less than 20
years, at least one component with a mass less than 20\% of the Sun's;
for details see \citet{1999ApJ...512..864H}), which includes red dwarf
binaries observed to measure high quality masses.  Presumed single
main sequence stars are shown as open circles and known close
multiples with combined photometry are shown with filled squares.  

The solid line represents identical photometric and trigonometric
distances.  Dotted lines trace distances differing by 20\% --- stars
above the outlined region are often subdwarfs; stars below are often
unresolved multiples because they are predicted photometrically to be
closer than they actually are.  In two cases, SCR 1845-6357 AB and LP
771-095 BC, the multiples are within the 20\% region, as is expected
because of the relatively large magnitude differences between the
components.  Additional close multiples in which secondaries
contribute minimal light to the system would not be revealed with this
method.  From left to right (increasing trigonometric distance), the
known close multiples recovered below the 20\% region are G 041-014
ABC, GJ 54 AB, GJ 2005 ABC, G 089-032 AB, SCR 0630-7643 AB, and GJ
2130 BC.  The three best candidates for previously unnoticed multiples
are shown as heavy open squares --- from left to right, GJ 555, ER 2,
and LTT 6933.  GJ 555 is a marginal case, while the latter two are
almost certainly close multiples with small flux ratios.

\subsection{The Big Picture}
\label{sec:disc.future}

To date, we have observed more than 400 stars for parallax during
CTIOPI.  This paper is the fifth in our series that provides accurate
parallaxes for nearby stellar systems --- one from Ianna's plate
parallax effort at Siding Spring Observatory and four from the ongoing
CCD effort at CTIO.  With the 20 new systems described here, we have
now reported high quality parallaxes for a total of 29 new RECONS
systems.  We anticipate that future papers will yield additional
nearby systems because the pool of stars nearer than 10 pc without
accurate parallaxes is by no means drained.

We emphasize that the stellar companions discussed here are not at
CTIOPI's detection threshold --- the ASPENS program will search for
substellar and possibly planetary mass companions.  Red and white
dwarf systems within 10 pc and south of $\delta$ $=$ 0, including all
but one (WT 460) of the new RECONS members discussed here, are being
observed intensely to reveal any possible long term astrometric
perturbations.  Initial results imply that we can detect companions
with masses as low as 10 M$_{Jup}$ for stars with good observational
coverage and strong reference star sets.

Each new system found within 10 pc is a member of the fundamental
sample of objects that is used to: (1) determine accurate luminosity
and mass functions in the solar neighborhood, (2) evaluate the
multiplicity of stellar populations, and (3) calculate the total
amount of mass found in stars and substellar objects.  Because of
their proximity, each system is a promising target for detailed
astrophysical studies such as astroseismology, variability
investigations, direct radii measurements, and metal abundance
evaluations.  Aside from astrophysical efforts, these systems have
great astrobiological potential because of their proximity.  They are
prime targets to be scrutinized as possible planetary hosts, and any
planets found can be explored as nearby harbors of life.

\section{Acknowledgments}

We would like to thank Georgia State University students who have
assisted in many facets of the RECONS effort to discover nearby stars,
including Jacob Bean, Misty Brown, Charlie Finch, Stephanie Ramos, and
Jennifer Winters.  We also gratefully acknowledge assistance in the
early stages of the CTIOPI effort from Claudio Anguita, Rafael Pujals,
Maria Teresa Ruiz and Pat Seitzer.  Nigel Hambly has been vital to our
SuperCOSMOS search, which revealed the three SCR stars here for which
we report parallaxes.  We also thank Hartmut Jahrei{\ss}, a referee
without peer for this paper, for his review.

Without the extensive observing support of Alberto Miranda, Edgardo
Cosgrove, Arturo Gomez and the staff at CTIO, the results presented
here would not be possible.  We are deeply indebted to NOAO for
providing us a long term observing program at CTIO via the NOAO
Surveys Project.  We also thank the continuing support of the members
of the SMARTS Consortium without whom many of the astrometric series
reported would not have been completed.

The early phase of CTIOPI was supported by the NASA/NSF Nearby Star
(NStars) Project through NASA Ames Research Center.  The RECONS team
at Georgia State University (GSU) is supported by NASA's Space
Interferometry Mission, the National Science Foundation (grant
AST-0507711), and GSU.  Ianna's efforts at the University of Virginia
have been supported by the National Science Foundation (grant
AST-9820711).  EC and RAM acknowledge support by the Fondo Nacional de
Investigaci\'on Cient\'ifica y Tecnol\'ogica (proyecto Fondecyt
No. 1010137), and by the Chilean Centro de Astrof\'isica FONDAP
(No. 15010003).  This work has used data products from the Two Micron
All Sky Survey, which is a joint project of the University of
Massachusetts and the Infrared Processing and Analysis Center at
California Institute of Technology funded by NASA and NSF.


\clearpage


\figcaption[fig1] {The nightly astrometric residuals in RA and DEC are
shown for the binary star system, GJ 54 AB, in the top two panels,
after fitting for parallax and proper motion.  The 1.1 year orbital
perturbation in the photocenter makes it difficult to fit the 1.0 year
period of the parallactic motion, resulting in relatively high RMS
residuals of 9.1 and 13.6 mas in RA and DEC, respectively.  Some
seasonal trends are seen in the data.  The much lower residuals for
the presumed single star, GJ 300, are shown in the bottom two panels,
with RMS values of only 3.6 mas in both RA and DEC.  For GJ 300, a 10
M$_{Jup}$ planet in a face-on circular orbit with a period of six
years would cause the photocenter to trace a circle with a semimajor
axis of 12 mas, clearly ruled out by the data.  The semimajor axis
expands to 19 mas for the same object in a 12 year orbit, which is
twice the time coverage of the astrometric series currently available.
For both systems, points with circles around them indicate nights on
which only one frame was taken, and representative errors have been
assigned.
\label{fig:henry1}}

\figcaption[fig2] {The RECONS sample of objects in systems with
trigonometric parallaxes larger than 100 mas and errors less than 10
mas is plotted on an HR Diagram, using $M_V$ vs. $V-K$.  Noteworthy
stars are labeled, including the four known subdwarfs (triangles), the
three $\sim$10 Myr old objects AU Mic and AT Mic A and B, and some of
the recently discovered members of the 10 pc sample.  Open points
represent objects known in 2000 or discovered recently by other
groups, and solid points represent RECONS discoveries.  The three L
dwarfs redder than $V-K$ $=$ 10 have all had distances measured since
2000.
\label{fig:henry2}}

\figcaption[fig3] {A comparison of photometric distances and
trigonometric distances is shown for objects within 20 pc with
parallaxes published in the four CTIOPI papers reporting parallaxes to
date.  Open points are presumed single main sequence stars.  Solid
squares are known multiples with combined photometry.  Three possible
unnoticed doubles are shown with open squares --- from left to right,
GJ 555, ER 2, and LTT 6933, all from \citet{2005AJ....129.1954J}, and
LTT 6933 confirmed in \citet{2005AJ....130..337C}.  The solid line
represents equal photometric and trigonometric distances, and the
dotted lines trace the region with distances differing by more than
20\%.
\label{fig:henry3}}


\begin{figure}
\begin{center}
\includegraphics[scale=0.60,angle=90]{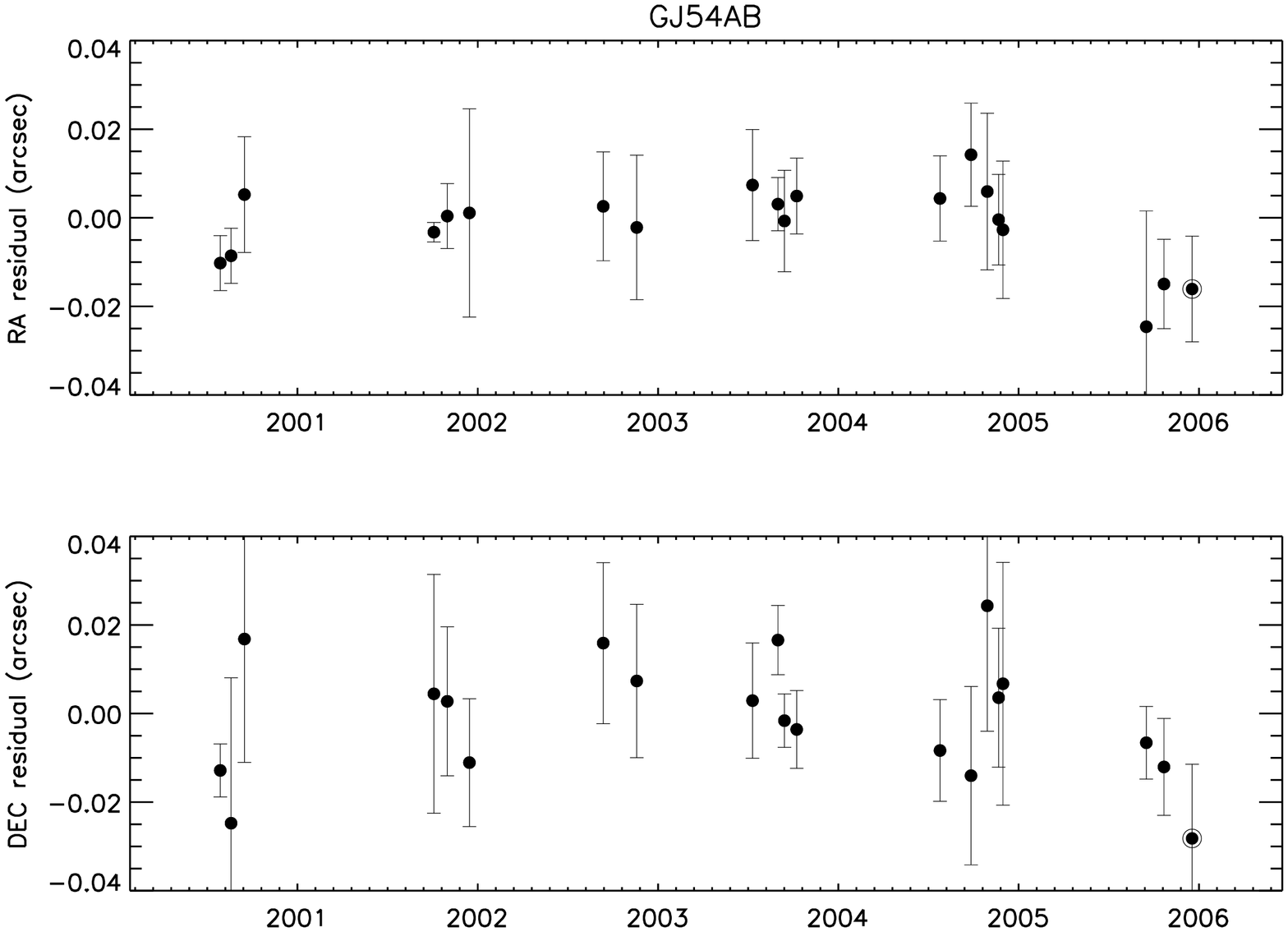}\\
\includegraphics[scale=0.60,angle=90]{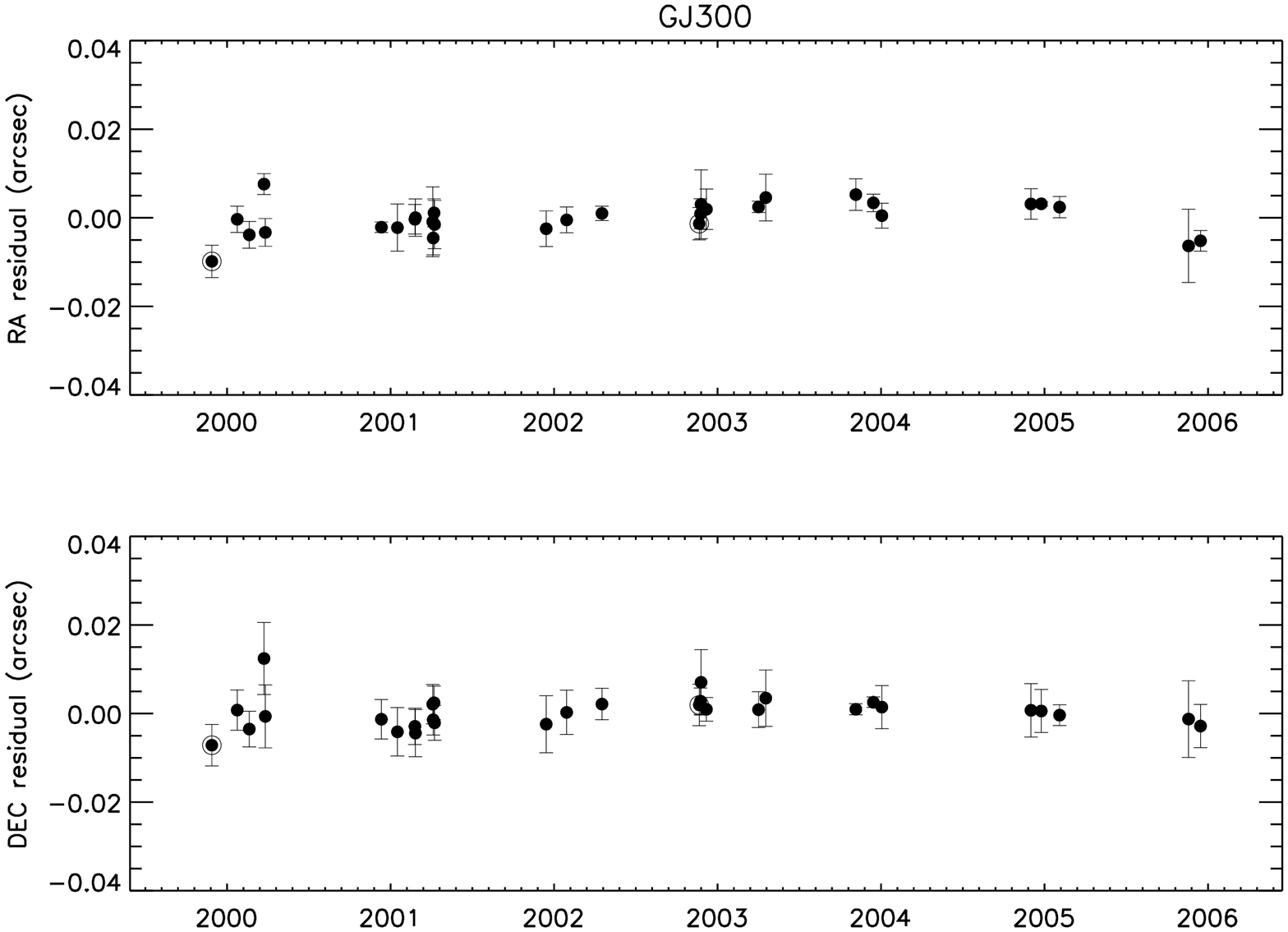}
\end{center}
\end{figure}

\clearpage

\begin{figure}
\includegraphics[scale=0.85,angle=00]{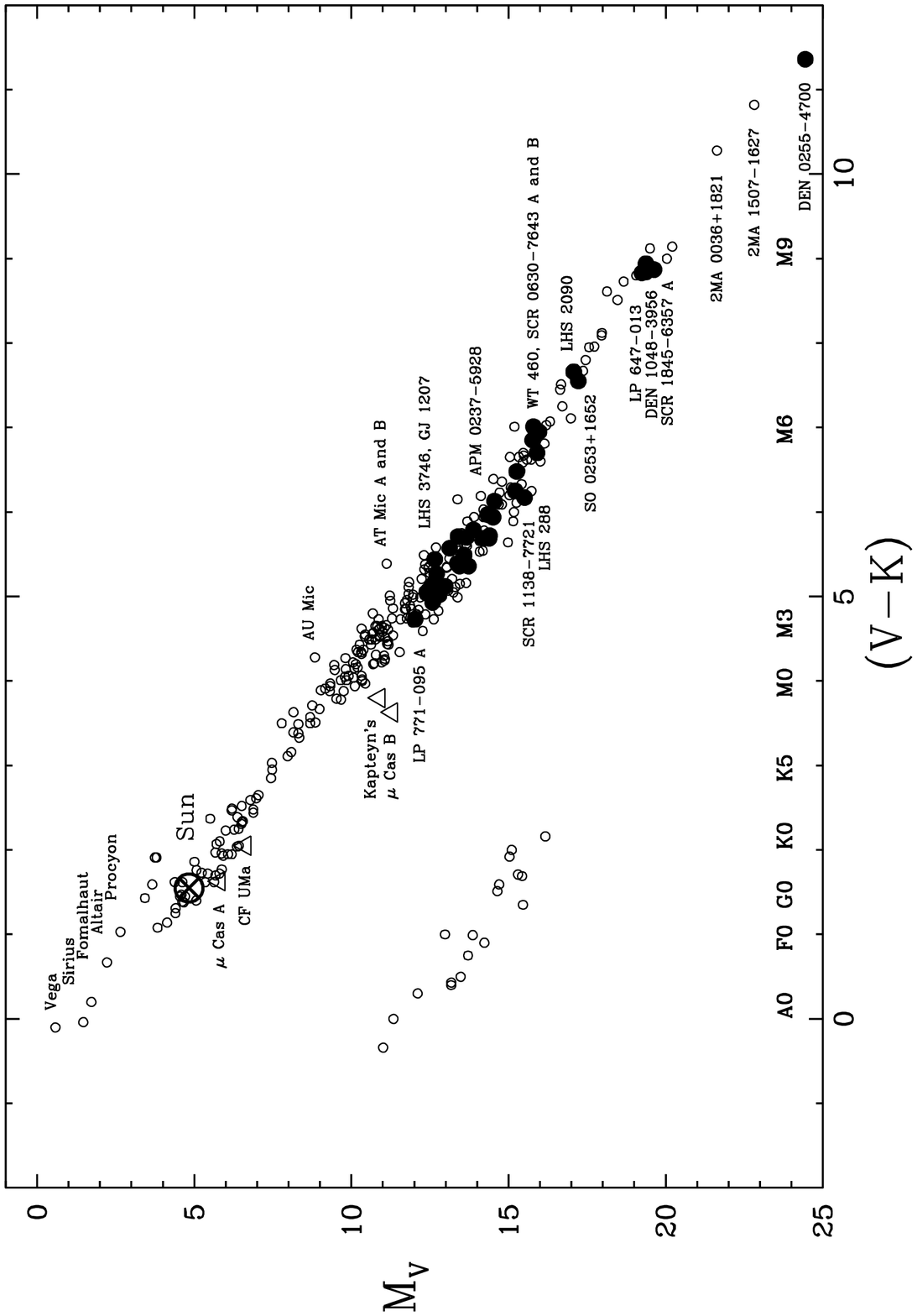}
\end{figure}

\clearpage

\begin{figure}
\includegraphics[scale=0.85,angle=00]{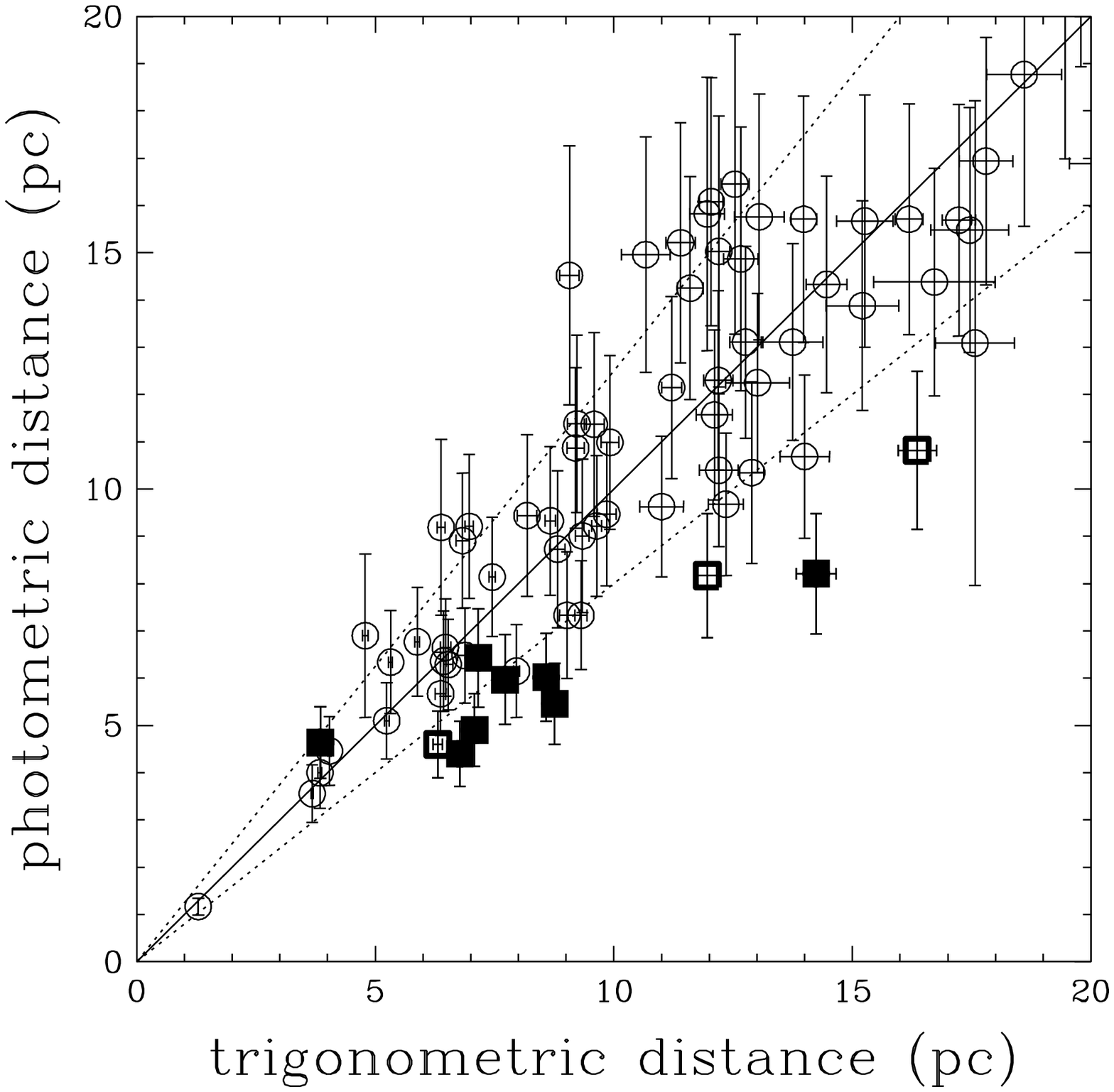}
\end{figure}

\clearpage


\voffset0pt{
\centering
\begin{deluxetable}{lcccl}
\tabletypesize{\scriptsize}
\tablecaption{Ground-Based Trigonometric Parallaxes for Stars within 10 Parsecs Since YPC.\tablenotemark{a}}
\setlength{\tabcolsep}{0.03in}
\label{tab:piground}
\tablewidth{0pt}

\tablehead{
	   \colhead{Reference}                       &
	   \colhead{Observing}                       &
 	   \colhead{\# Total}                        &
 	   \colhead{\# New 10 pc}                    &
 	   \colhead{New 10 pc}                       \\
                                                        
           \colhead{}                                &
           \colhead{Method}                          &
           \colhead{Parallaxes\tablenotemark{b}}     &
           \colhead{Systems\tablenotemark{c}}        &
           \colhead{Systems\tablenotemark{c}}        }
\startdata
\multicolumn{5}{c}{Other Parallax Programs}                                                                    \\
\hline							      		      	      	 	    	     	      	      	      	      	      	      
                            &                   &     &     &                                                  \\
\citet{1996MNRAS.281..644T} & optical CCD       &  13 &   1 & LP 944-020 ($=$ BRI 0337-3535)                   \\
\citet{2002AJ....124.1170D} & optical CCD       &  28 &   2 & 2MA 0036+1821, 2MA 1507-1627                     \\
\citet{2003AJ....125..354R} & optical CCD       &   1 &   1 & 2MA 1835+3259                                    \\
\citet{2004AJ....127.2948V} & IR array          &  40 &   3 & 2MA 0415-0935, 2MA 0727+1710, 2MA 0937+2931      \\
                            &                   &     &     &                                                  \\
TOTAL (other programs)      &                   &  82 &   7 &                                                  \\
                            &                   &     &     &                                                  \\
\hline						      	  		      	      	 	    	     	      	      	      	      	      	      
\multicolumn{5}{c}{RECONS Efforts}                                                                             \\
\hline						      	  		      	      	 	    	     	      	      	      	      	      	      
                            &                   &     &     &                                                  \\
\citet{1997AJ....114..388H} & phot plates       &   1 &   1 & GJ 1061                                          \\
\citet{2005AJ....129.1954J} & optical CCD       &  46 &   5 & GJ 754, GJ 1068, GJ 1123, GJ 1128, DEN 1048-3956\tablenotemark{d} \\
\citet{2005AJ....130..337C} & optical CCD       &  31 &   2 & GJ 2005 ABC, LP 647-013                          \\
Costa et al.~(2006, AJ in press) & optical CCD  &  31 &   1 & DEN 0255-4700                                    \\
this paper                  & optical CCD       &  27 &  20 & see Table 3                                      \\
                            &                   &     &     &                                                  \\
TOTAL (RECONS)              &                   & 136 &  29 &                                                  \\
                            &                   &     &     &                                                  \\
\enddata

\tablenotetext{a}{papers listed describe new systems within 10 pc having parallax errors less than 10 mas}
\tablenotetext{b}{reference star parallaxes not included in counts}
\tablenotetext{c}{error in weighted mean of all $\pi_{trig}$ must be less than 10 mas}
\tablenotetext{d}{also reported in \citet{2005AJ....130..337C} from CTIOPI at the 1.5 m}
\end{deluxetable}
}


\voffset000pt{
\begin{deluxetable}{llr@{.}lr@{.}lr@{.}lcccccr@{.}lr@{.}lr@{.}lccc}
\rotate
\setlength{\tabcolsep}{0.03in}
\label{tab:photspec}
\tablewidth{0pt}
\tabletypesize{\scriptsize}
\tablecaption{Photometric and Spectroscopic Results.}
\tablehead{\colhead{Name1}               &
           \colhead{Name2}               &
           \multicolumn{2}{c}{$V_{J}$}   &
           \multicolumn{2}{c}{$R_{KC}$}  &
           \multicolumn{2}{c}{$I_{KC}$}  &
           \colhead{\# nts}              &
           \colhead{$\pi$ filter}        &
           \colhead{$\sigma$ (mag)}      &
           \colhead{\# nts}              &
           \colhead{\# frm}              &
           \multicolumn{2}{c}{$J$}       &
           \multicolumn{2}{c}{$H$}       &
           \multicolumn{2}{c}{$K_{s}$}   &
	   \colhead{d$_{phot}$}          &
	   \colhead{\# rel}              &
	   \colhead{SpType}              \\

	   \colhead{(1)}                 &
	   \colhead{(2)}                 &
           \multicolumn{2}{c}{(3)}       &
           \multicolumn{2}{c}{(4)}       &
           \multicolumn{2}{c}{(5)}       &
	   \colhead{(6)}                 &
           \colhead{(7)}                 &
           \colhead{(8)}                 &
           \colhead{(9)}                 &
           \colhead{(10)}                &
           \multicolumn{2}{c}{(11)}      &
           \multicolumn{2}{c}{(12)}      &
           \multicolumn{2}{c}{(13)}      &
	   \colhead{(14)}                &
	   \colhead{(15)}                &
	   \colhead{(16)}                }
\startdata
\multicolumn{22}{c}{New 10 Parsec Members}\\				      		 	    
\hline							      		      	      	 	    
LHS 1302         & G 159-003       &  14&49  &  13&00  &  11&17  &  5 &  R & 0.021v & 26 & 141 & 9&41  &    8&84  &    8&55  &  10.99 $\pm$ 1.84 & 12 & M5.0 V  \\
APMPM J0237-5928 &                 &  14&47  &  12&96  &  11&08  &  5 &  R & 0.013v & 27 & 166 & 9&28  &    8&70  &    8&34  &   9.22 $\pm$ 1.49 & 12 & M5.0 V  \\
SO 0253+1652     &                 &  15&14  &  13&03  &  10&65  &  3 &  I & 0.006  & 16 &  95 & 8&39  &    7&88  &    7&59  &   4.00 $\pm$ 0.75 & 12 & M7.0 V  \\
LP 771-095 A     &                 &  11&22  &  10&07  &   8&66  &  4 &  V & 0.012  & 22 & 109 & 7&29  &    6&77  &    6&50  &   8.91 $\pm$ 1.43 & 12 & M3.0 V  \\
LP 771-095 BC    & LP 771-096      &  11&37J &  10&13J &   8&58J &  4 &  V & 0.043  & 21 &  98 & 7&11J &    6&56J &    6&29J &   6.43 $\pm$ 1.04 & 12 & M3.5 VJ \\
LHS 1610         & G 006-039       &  13&85  &  12&42  &  10&66  &  5 &  V & 0.024v & 23 & 128 & 8&93  &    8&38  &    8&05  &   9.47 $\pm$ 1.52 & 12 & M4.5 V  \\
LHS 1723         &                 &  12&22  &  10&87  &   9&18  &  5 &  V & 0.020v & 27 & 207 & 7&62  &    7&07  &    6&74  &   6.34 $\pm$ 1.09 & 12 & M4.5 V  \\
G 099-049        & LTT 17897       &  11&31  &  10&04  &   8&42  &  4 &  V & 0.012  & 23 & 145 & 6&91  &    6&31  &    6&04  &   5.10 $\pm$ 0.81 & 12 & M4.5 V  \\
SCR 0630-7643 AB &                 &  14&82J &  13&08J &  11&00J &  4 &  I & 0.005  & 12 &  69 & 8&89J &    8&28J &    7&92J &   5.46 $\pm$ 0.86 & 12 & M6.0 VJ \\
G 089-032 AB     & LTT 17993       &  13&25J &  11&81J &   9&97J &  4 &  R & 0.009  & 32 & 215 & 8&18J &    7&61J &    7&28J &   6.02 $\pm$ 0.93 & 12 & M5.0 VJ \\
GJ 300           & LHS 1989        &  12&15  &  10&85  &   9&22  &  3 &  V & 0.016v & 31 & 191 & 7&60  &    6&96  &    6&71  &   6.15 $\pm$ 0.98 & 12 & M3.5 V  \\
G 041-014 ABC    &                 &  10&92J &   9&67J &   8&05J &  3 &  V & 0.009  & 22 & 160 & 6&51J &    5&97J &    5&69J &   4.40 $\pm$ 0.69 & 12 & M4.5 VJ \\
LHS 2090         &                 &  16&10  &  14&11  &  11&84  &  2 &  I & 0.007  & 15 &  71 & 9&44  &    8&84  &    8&44  &   5.67 $\pm$ 0.88 & 12 & M6.0 V  \\
LHS 2206         & G 042-024       &  14&02  &  12&63  &  10&85  &  3 &  R & 0.011  & 20 & 118 & 9&21  &    8&60  &    8&33  &  11.38 $\pm$ 1.88 & 12 & M4.5 V  \\
LHS 288          &                 &  13&90  &  12&31  &  10&27  &  3 &  R & 0.007  & 13 &  68 & 8&49  &    8&05  &    7&73  &   6.90 $\pm$ 1.73 & 12 & M5.0 V  \\
SCR 1138-7721    &                 &  14&78  &  13&20  &  11&24  &  4 &  I & 0.004  & 12 &  59 & 9&40  &    8&89  &    8&52  &   9.44 $\pm$ 1.71 & 12 & M5.5 V  \\
LHS 337          &                 &  12&75  &  11&44  &   9&74  &  3 &  R & 0.006  & 10 &  50 & 8&17  &    7&76  &    7&39  &   9.19 $\pm$ 1.86 & 12 & M4.0 V  \\
WT 460           &                 &  15&63  &  13&90  &  11&78  &  3 &  I & 0.008  & 28 & 151 & 9&67  &    9&04  &    8&62  &   7.33 $\pm$ 1.15 & 12 & M5.5 V  \\
GJ 1207          & LHS 3255        &  12&25  &  11&00  &   9&43  &  5 &  V & 0.263v & 27 & 124 & 7&97  &    7&44  &    7&12  &   9.33 $\pm$ 1.57 & 12 & M3.5 V  \\
SCR 1845-6357 AB &                 &  17&40J &  15&00J &  12&46J &  5 &  I & 0.004  & 18 & 117 & 9&54J &    8&97J &    8&51J &   4.64 $\pm$ 0.76 & 10 & M8.5 VJ \\
LHS 3746         &                 &  11&76  &  10&56  &   9&04  &  4 &  V & 0.013  & 33 & 213 & 7&60  &    7&02  &    6&72  &   8.14 $\pm$ 1.26 & 12 & M3.0 V  \\
\hline
\multicolumn{22}{c}{Known 10 Parsec Members}\\
\hline
GJ 54 AB         & LHS 1208        &   9&82J &   8&70J &   7&32J &  5 &  V & 0.015  & 20 & 149 & 6&00J &    5&41J &    5&13J &   4.90 $\pm$ 0.77 & 12 & M3.0 VJ \\
GJ 1061          & LHS 1565        &  13&09  &  11&45  &   9&46  &  5 &  R & 0.010  & 27 & 186 & 7&52  &    7&02  &    6&61  &   3.56 $\pm$ 0.61 & 12 & M5.5 V  \\
\hline
\multicolumn{22}{c}{Not 10 Parsec Members}\\
\hline
GJ 1001 A        & LHS 102         &  12&86  &  11&64  &  10&10  &  2 &  R & 0.010  & 14 &  61 & 8&60  &    8&04  &    7&74  &  12.25 $\pm$ 1.89 & 12 & M3.5 V  \\
GJ 633           & LHS 3233        &  12&66  &  11&55  &  10&18  &  4 &  V & 0.007  & 14 &  64 & 8&89  &    8&31  &    8&05  &  19.70 $\pm$ 3.11 & 12 & M2.5 V  \\
GJ 2130 A        & CD -32 13297    &  10&50  &   9&49  &   8&34  &  4 &  V & 0.009  & 12 &  71 & 7&11  &    6&53  &    6&25  &  10.69 $\pm$ 1.73 & 12 & M1.5 V  \\
GJ 2130 BC       & CD -32 13298    &  11&51J &  10&34J &   8&86J &  4 &  V & 0.010  & 13 & 113 & 7&46J &    6&86J &    6&59J &   8.21 $\pm$ 1.27 & 12 & M3.0 VJ \\
\enddata

\tablecomments{J: photometry or spectral type is for more than one object, i.e. joint}
\end{deluxetable}
}


\clearpage

\thispagestyle{empty}

\voffset000pt{
\begin{deluxetable}{lccccccccrrrrrrc}
\rotate
\setlength{\tabcolsep}{0.03in}
\label{tab:pictiopi}
\tablewidth{0pt}
\tabletypesize{\scriptsize}
\tablecaption{Astrometric Results.}
\tablehead{\colhead{}                    &
	   \colhead{R.A.}                &
 	   \colhead{Decl.}               &
 	   \colhead{}                    &
	   \colhead{}                    &
	   \colhead{}                    &
	   \colhead{}                    &
	   \colhead{}                    &
	   \colhead{}                    &
	   \colhead{$\pi$(rel)}          &
	   \colhead{$\pi$(corr)}         &
	   \colhead{$\pi$(abs)}          &
	   \colhead{$\mu$}               &
	   \colhead{P.A.}                &
	   \colhead{$V_{tan}$}           \\

           \colhead{Name}                &
	   \colhead{J2000.0}             &
 	   \colhead{J2000.0}             &
 	   \colhead{Filter}              &
	   \colhead{N$_{sea}$}           &
	   \colhead{N$_{frm}$}           &
	   \colhead{Coverage}            &
	   \colhead{Years}               &
 	   \colhead{N$_{ref}$}           &
	   \colhead{mas}                 &
	   \colhead{mas}                 &
	   \colhead{mas}                 &
	   \colhead{mas yr$^{-1}$}       &
	   \colhead{deg}                 &
	   \colhead{km sec$^{-1}$}       &
	   \colhead{notes}               \\

           \colhead{(1)}                 &
           \colhead{(2)}                 &
           \colhead{(3)}                 &
           \colhead{(4)}                 &
           \colhead{(5)}                 &
           \colhead{(6)}                 &
           \colhead{(7)}                 &
           \colhead{(8)}                 &
           \colhead{(9)}                 &
           \colhead{(10)}                &
           \colhead{(11)}                &
           \colhead{(12)}                &
           \colhead{(13)}                &
           \colhead{(14)}                &
           \colhead{(15)}                &
           \colhead{(16)}                }
\startdata
\multicolumn{16}{c}{New 10 Parsec Members}\\
\hline
LHS 1302         & 01 51 04.09 & $-$06 07 05.1 & R & 7c & 141 & 1999.71$-$2005.96 & 6.25 &  6 & 100.14$\pm$1.89 & 0.64$\pm$0.06 & 100.78$\pm$1.89 &  597.1$\pm$0.9 & 115.5$\pm$0.17 &  28.1 &                   \\ 
APMPM J0237-5928 & 02 36 32.46 & $-$59 28 05.7 & R & 7c & 160 & 1999.64$-$2005.96 & 6.32 &  6 & 102.38$\pm$1.11 & 1.34$\pm$0.11 & 103.72$\pm$1.12 &  724.9$\pm$0.5 &  52.2$\pm$0.08 &  33.1 &                   \\ 
SO 0253+1652     & 02 53 00.89 & $+$16 52 52.7 & I & 3c &  95 & 2003.53$-$2005.88 & 2.35 &  6 & 258.48$\pm$2.68 & 2.15$\pm$0.21 & 260.63$\pm$2.69 & 5106.8$\pm$3.1 & 138.2$\pm$0.07 &  92.9 & \tablenotemark{a} \\  
LP 771-095 A     & 03 01 51.39 & $-$16 35 36.0 & V & 6c & 109 & 1999.64$-$2005.96 & 6.33 &  5 & 145.44$\pm$2.92 & 0.95$\pm$0.14 & 146.39$\pm$2.92 &  477.0$\pm$1.4 & 234.4$\pm$0.34 &  15.4 & \tablenotemark{b} \\
LP 771-095 BC    & 03 01 51.04 & $-$16 35 31.0 & V & 6c &  99 & 1999.64$-$2005.96 & 6.15 &  5 & 138.75$\pm$4.99 & 0.95$\pm$0.14 & 139.70$\pm$4.99 &  479.4$\pm$2.5 & 235.5$\pm$0.59 &  16.3 &                   \\ 
LHS 1610         & 03 52 41.76 & $+$17 01 04.3 & V & 7c & 128 & 1999.71$-$2005.96 & 6.25 &  6 & 100.24$\pm$2.07 & 1.33$\pm$0.12 & 101.57$\pm$2.07 &  767.0$\pm$1.0 & 146.1$\pm$0.15 &  35.8 & \tablenotemark{c} \\   
LHS 1723         & 05 01 57.43 & $-$06 56 46.5 & V & 7c & 207 & 1999.81$-$2005.95 & 6.14 & 11 & 186.20$\pm$1.25 & 1.72$\pm$0.19 & 187.92$\pm$1.26 &  769.4$\pm$0.7 & 226.9$\pm$0.10 &  19.4 &                   \\ 
G 099-049        & 06 00 03.52 & $+$02 42 23.6 & V & 7c & 145 & 1999.91$-$2005.96 & 6.06 &  7 & 189.43$\pm$1.82 & 1.50$\pm$0.50 & 190.93$\pm$1.89 &  312.5$\pm$0.8 &  97.5$\pm$0.24 &   7.8 & \tablenotemark{d} \\
SCR 0630-7643 AB & 06 30 46.61 & $-$76 43 09.0 & I & 3c &  69 & 2003.94$-$2005.97 & 2.03 &  8 & 112.07$\pm$1.84 & 2.09$\pm$0.19 & 114.16$\pm$1.85 &  455.7$\pm$3.0 & 356.9$\pm$0.53 &  18.9 &                   \\ 
G 089-032 AB     & 07 36 25.13 & $+$07 04 43.1 & R & 7c & 216 & 1999.91$-$2005.95 & 6.05 & 13 & 115.67$\pm$0.97 & 0.93$\pm$0.08 & 116.60$\pm$0.97 &  396.9$\pm$0.5 & 143.0$\pm$0.14 &  16.1 &                   \\ 
GJ 300           & 08 12 40.88 & $-$21 33 06.8 & V & 7c & 191 & 1999.91$-$2005.95 & 6.05 &  8 & 122.40$\pm$0.90 & 3.20$\pm$0.37 & 125.60$\pm$0.97 &  698.8$\pm$0.5 & 178.7$\pm$0.06 &  26.4 & \tablenotemark{e} \\
G 041-014 ABC    & 08 58 56.33 & $+$08 28 26.0 & V & 7c & 159 & 1999.97$-$2005.96 & 5.99 &  5 & 145.40$\pm$1.97 & 2.26$\pm$0.22 & 147.66$\pm$1.98 &  502.7$\pm$0.9 & 130.0$\pm$0.20 &  16.1 &                   \\ 
LHS 2090         & 09 00 23.55 & $+$21 50 04.8 & I & 4c &  71 & 2002.28$-$2005.20 & 2.92 &  9 & 155.98$\pm$2.67 & 0.89$\pm$0.06 & 156.87$\pm$2.67 &  773.9$\pm$2.2 & 221.2$\pm$0.33 &  23.4 &                   \\ 
LHS 2206         & 09 53 55.19 & $+$20 56 46.8 & R & 7c & 118 & 2000.06$-$2005.97 & 5.91 &  6 & 107.89$\pm$2.30 & 0.50$\pm$0.03 & 108.39$\pm$2.30 &  522.6$\pm$1.0 & 321.1$\pm$0.22 &  22.9 &                   \\ 
LHS 288          & 10 44 21.23 & $-$61 12 35.6 & R & 5s &  63 & 2000.06$-$2005.13 & 5.07 & 10 & 207.73$\pm$2.73 & 1.22$\pm$0.16 & 208.95$\pm$2.73 & 1642.9$\pm$1.2 & 347.7$\pm$0.07 &  37.3 & \tablenotemark{f} \\
SCR 1138-7721    & 11 38 16.76 & $-$77 21 48.5 & I & 3s &  59 & 2003.23$-$2005.33 & 2.09 & 11 & 120.25$\pm$2.91 & 2.02$\pm$0.28 & 122.27$\pm$2.92 & 2147.6$\pm$5.1 & 286.9$\pm$0.25 &  83.3 &                   \\ 
LHS 337          & 12 38 49.10 & $-$38 22 53.8 & R & 4s &  50 & 2002.28$-$2005.47 & 3.19 & 14 & 155.37$\pm$1.99 & 1.41$\pm$0.14 & 156.78$\pm$1.99 & 1464.3$\pm$1.8 & 206.4$\pm$0.13 &  44.3 &                   \\ 
WT 460           & 14 11 59.94 & $-$41 32 21.3 & I & 6c & 152 & 2000.14$-$2005.57 & 5.43 & 10 & 105.58$\pm$1.51 & 1.83$\pm$0.13 & 107.41$\pm$1.52 &  690.4$\pm$0.9 & 260.0$\pm$0.12 &  31.0 &                   \\ 
GJ 1207          & 16 57 05.73 & $-$04 20 56.3 & V & 7c & 124 & 1999.62$-$2005.71 & 6.09 & 10 & 113.36$\pm$1.44 & 2.03$\pm$0.44 & 115.39$\pm$1.51 &  608.5$\pm$0.8 & 127.1$\pm$0.15 &  25.0 & \tablenotemark{g} \\
SCR 1845-6357 AB & 18 45 05.26 & $-$63 57 47.8 & I & 3c & 117 & 2003.24$-$2005.81 & 2.57 & 12 & 258.29$\pm$1.11 & 1.16$\pm$0.08 & 259.45$\pm$1.11 & 2664.4$\pm$1.7 &  76.6$\pm$0.06 &  48.7 & \tablenotemark{h} \\
LHS 3746         & 22 02 29.39 & $-$37 04 51.3 & V & 7c & 213 & 1999.71$-$2005.88 & 6.17 &  7 & 133.14$\pm$1.20 & 1.15$\pm$0.11 & 134.29$\pm$1.31 &  820.7$\pm$0.8 & 105.1$\pm$0.09 &  29.0 &                   \\ 
\hline						     	      			      	      			  		  		   		    			     		   
\multicolumn{16}{c}{Known 10 Parsec Members} \\
\hline						     	      			      	      			  		  		   		    			     		   
GJ 54 AB         & 01 10 22.90 & $-$67 26 41.9 & V & 6c & 149 & 2000.57$-$2005.96 & 5.39 &  8 & 139.28$\pm$3.32 & 1.92$\pm$0.62 & 141.20$\pm$3.38 &  682.5$\pm$2.1 &  33.1$\pm$0.34 &  22.9 & \tablenotemark{i} \\ 
GJ 1061          & 03 35 59.71 & $-$44 30 45.4 & R & 7c & 186 & 1999.62$-$2005.95 & 6.33 &  7 & 270.98$\pm$1.34 & 0.94$\pm$0.08 & 271.92$\pm$1.34 &  826.3$\pm$0.8 & 117.7$\pm$0.10 &  14.4 & \tablenotemark{j} \\ 
\hline						     	      			      	      			  		  		   		    			     		   
\multicolumn{16}{c}{Not 10 Parsec Members} \\
\hline						     	      			      	      			  		  		   		    			     		   
GJ 1001 A        & 00 04 36.46 & $-$40 44 02.7 & R & 5s &  61 & 1999.64$-$2005.95 & 6.32 &  6 &  75.83$\pm$3.97 & 1.03$\pm$0.07 &  76.86$\pm$3.97 & 1627.0$\pm$1.8 & 156.7$\pm$0.12 & 100.3 & \tablenotemark{k} \\
GJ 633           & 16 40 45.26 & $-$45 59 59.3 & V & 6s &  62 & 1999.64$-$2005.71 & 6.07 & 10 &  41.78$\pm$1.75 & 2.71$\pm$0.57 &  44.49$\pm$1.84 &  528.4$\pm$1.4 & 136.6$\pm$0.30 &  56.3 & \tablenotemark{l} \\
GJ 2130 A        & 17 46 12.75 & $-$32 06 09.3 & V & 5s &  71 & 1999.64$-$2005.71 & 6.07 &  8 &  68.85$\pm$2.55 & 2.60$\pm$0.68 &  71.45$\pm$2.64 &  277.8$\pm$2.4 & 196.2$\pm$0.87 &  18.4 & \tablenotemark{m} \\ 
GJ 2130 BC       & 17 46 14.42 & $-$32 06 08.5 & V & 5s & 113 & 1999.64$-$2005.71 & 6.07 &  8 &  67.63$\pm$1.93 & 2.60$\pm$0.68 &  70.23$\pm$2.05 &  277.4$\pm$1.8 & 196.4$\pm$0.66 &  18.7 &                   \\ 
\enddata

\vfil\eject

\tablecomments{coordinates are epoch and equinox 2000.0; each target's coordinates were extracted from 2MASS and then transformed to epoch 2000.0 using the proper motions and position angles listed here}
\tablenotetext{a}{parallax of 410 $\pm$ 90 mas in \citet{2003ApJ...589L..51T}}
\tablenotetext{b}{parallax of 92.97 $\pm$ 38.04 mas in HIP}
\tablenotetext{c}{parallax of 70.0 $\pm$ 13.8 mas in YPC}
\tablenotetext{d}{parallax of 186.2 $\pm$ 10.1 mas in YPC}
\tablenotetext{e}{parallax of 169.9 $\pm$ 15.0 mas in YPC}
\tablenotetext{f}{parallax of 222.5 $\pm$ 11.3 mas in YPC}
\tablenotetext{g}{parallax of 104.4 $\pm$ 13.6 mas in YPC}
\tablenotetext{h}{parallax of 282 $\pm$ 23 mas in \citet{2005AJ....129..409D}}
\tablenotetext{i}{parallaxes of 120.5 $\pm$ 10.1 mas in YPC and 122.86 $\pm$ 7.53 mas in HIP}
\tablenotetext{j}{parallax of 273.4 $\pm$ 5.2 mas in \citet{1997AJ....114..388H}}
\tablenotetext{k}{parallax of 104.7 $\pm$ 11.4 mas in YPC}
\tablenotetext{l}{parallax of 104.0 $\pm$ 13.7 mas in YPC}
\tablenotetext{m}{parallax of 161.77 $\pm$ 11.29 mas in HIP}
\end{deluxetable}
}

\end{document}